# Neutron diffraction strain tomography: 2D axisymmetric sample geometry

V. Luzin[1,2], A. Gregg[2], J. Hendriks[2], C. Wensrich[2]

[1] Australian Nuclear Science and Technology Organization (ANSTO), Lucas Heights, NSW 2234, Australia
[2] School of Engineering, The University of Newcastle, Callaghan, NSW 2308, Australia

**Abstract**

We consider a method for neutron residual strain/stress analysis combining features of transmission strain tomography (e.g. through Bragg's edge imaging) and neutron diffraction techniques. Analogous to transmission strain tomography, our approach observes integrals of strain along the path of a beam, however, rather than transmitted beams, we observe a diffracted beam. The practical feasibility of "neutron diffraction tomography" is demonstrated in case of a 2D axisymmetric sample where the reconstruction of the stress field is determined from the measurement of a strain profile. The reconstruction is compared to a set of measurements made using a traditional neutron stress scanning technique in order to validate the new method. Similar strain accuracy was achieved by the new technique within a reasonable measurement time. The practical and theoretical challenges regarding the technique are also discussed.

## 1. Introduction

In recent years, an interest in the topic of neutron strain tomography has risen, as confirmed by an increase in relevant publications [1-5]. It appears that this progress is related to technological advances in two area of science related to neutron scattering;

The first is a steady progress in the development of modern high-flux spallation neutron sources and associated instruments. Two prominent examples are SNS in the USA and J-PARC in Japan, which both started user operation around 2007. Two more neutron spallation sources are scheduled to begin operation in the near future; one in China (CSNS) and one in Europe (ESS).

The second reason is the availability of high spatial and time resolution neutron detectors with excellent efficiency [6, 7]. First developed for the purpose of space exploration by the Space Science Laboratory at the University of Berkley (USA), the MultiChannel Plate (MCP) detector has found its way towards more traditional applications in neutron radiography, tomography and imaging using neutron transmission spectrometry [8, 9]. In comparison to the first generation of energy resolved imaging detectors [Insert ref for ENGINX detector], the fast readout electronics (>100MHz), nanosecond temporal resolution, 55µm resolution and 28x28 mm$^2$ field-of-view of the MCP detector represented a major step forward that led to a wide range of practical applications [10].

With the current focus on transmission spectra using imaging beamlines, the usefulness of the diffraction signal can tend to be overlooked or be considered a by-product that is ignored. Traditional diffraction-based neutron strain measurement relies on the use of neutron optics (slits and radial collimators) to form a small gauge volume within the sample from which the strain can be measured (see Figure 1a). In contrast, imaging experiments fully immerse the sample in the incident beam and therefore the same concept of measuring spatially resolved strain seemingly cannot be applied.

Here we consider a technique where the diffracted signal can be potentially used simultaneously within a transmission strain tomography setup when a sample of axisymmetric symmetry is fully irradiated. The main idea is as follows:

If the sample is completely illuminated, and a vertical secondary slit is restricted to a certain resolution (i.e. slit size), the measured diffraction pattern originates from a gauge volume consisting of a vertical slice through the sample, and the measured signal is a bulk average over this slice (see Figure 1b). By sweeping the sample past the secondary slit, a profile of integrated strain measurements analogous to a transmission strain profile can be measured. Although similar in terms of the principle of a geometrical "projection", in terms of the measured strain, the scattering vector is not in the direction of the projection, rather it is determined by the scattering angle.

As with transmission strain profiles, this can form a natural tomography problem from which the strain distribution within the sample could be reconstructed. Since this approach combines both the tomographic principle and diffraction, in short this method is called henceforth "neutron diffraction tomography".

The considered method is somewhat similar to the method suggested by Korsunsky [11] for the case of high-energy synchrotron radiation experiments which inevitably involve elongated gauge volumes. Two distinct differences are apparent;

First, due to the small scattering angles associated with high-energy synchrotron radiation, the measurement direction is almost perpendicular to the incident beam. In the case of a neutron diffraction setup, the diffraction angle is much larger (typically 90°) and able to be varied over a wide range.

Second, in an axisymmetric system, it appears that the synchrotron setup is more convenient for measurement of the axial component (although an attempt to measure the radial component was also mentioned), while in the neutron diffraction setup, radial and hoop components are much easier to measure.

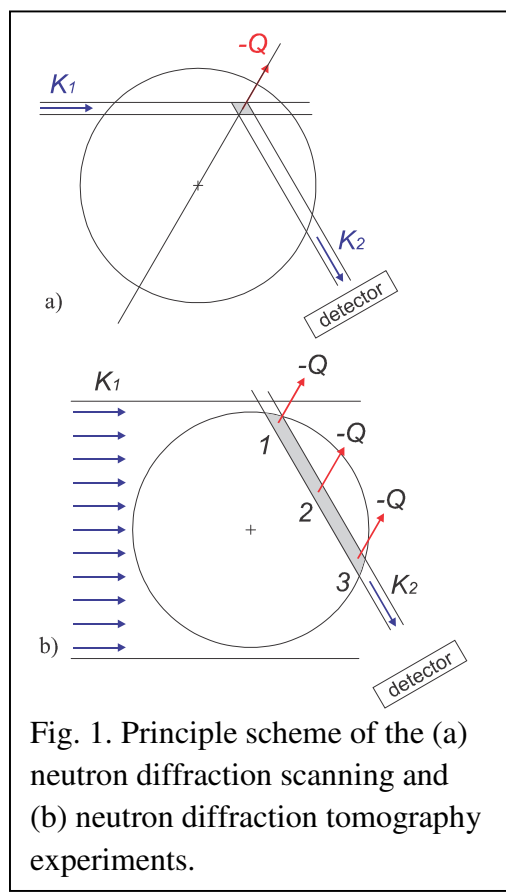

Fig. 1. Principle scheme of the (a) neutron diffraction scanning and (b) neutron diffraction tomography experiments.

To consider this a viable technique there are theoretical and practical issues to be resolved. Before examining the possibility for stain field reconstruction (i.e. the solution of the inverse tomographic problem), some technical questions must be answered. How is the strain field convoluted into the measured signal? Is this technique practical in terms of experimental procedure, measurement time, accuracy? Is it going to have acceptable and practically useful resolution?

In this paper we consider these challenges in the case of axisymmetric sample; the situation that was historically considered first during the development of neutron transmission tomography [11-13]. Due to the sample symmetry, multiple projections are not required which tremendously simplifies both the experiment and the reconstruction algorithm. At the same time it allows the examination of the underlying principles and demonstrates its practicality and feasibility at least for some sample types.

## 2. Theoretical background for axisymmetric geometry

We consider an axisymmetric object of radius R. For practical reasons, two cases can be considered; plane-stress and plane-stress conditions.

In a traditional strain scanning neutron diffraction experiment, strain measurements are done through collecting a diffraction pattern from a localised gauge volume formed by neutron optic elements (i.e. slits or radial collimators). This gauge volume should usually be smaller than the scale of the stress distribution to be resolved (Fig. 1a). In this case, for each sample orientation the localised gauge volume is characterised by a single scattering vector **Q** and a typical measurement protocol involves moving the gauge volume across sample and/or sample rotation to measure strain profiles for multiple strain components (directions in the sample). For example, in Fig. 1a a measurement scheme for the radial component of an axisymmetric sample is illustrated.

Alternatively, the diffraction experiment can be set up in an imaging sense where the sample is fully immersed in the incident neutron beam (Fig. 1b). In this case, the gauge volume is not well-localised, but is extended to the full dimension of the sample as defined by a secondary slit system (Fig. 1b). As a consequence, different parts of the long gauge volume are associated with different radial locations as well as different measurement directions in the polar coordinate system. To illustrate this fact, in Fig. 1b three locations are considered; in location 1 the measurement direction (direction of the scattering vector) is almost radial, in location 3 the measurement direction is approximately hoop, and for location 2 it is somewhere inbetween (i.e. a linear combination). Thus, the integration and averaging along the ray is not only over spatial locations, but also over different measurement directions. Although this feature is similar to the neutron transmission tomography experiment [5], the major difference is that the scattering vector is not parallel to the neutron beam direction, but determined by the scattering angle. A possible measurement protocol in this case would be to move the sample across the secondary slits so that all possible cross-sections are measured. It might also involve measurement at different Bragg angles through changing the scattering angle through changing reflections or changing the neutron beam wavelength.

The formal analysis of the measured signal in the described set-up involves evaluation of the integral along the line that correspond to the secondary beam and positioned at distance $\rho$ from the center of an axisymmetric sample of radius $R$:

$$\langle \varepsilon_q(\rho) \rangle = \frac{1}{2L}\int_{-L}^{+L} \varepsilon_q(s)ds = \frac{1}{2L}\int_{-L}^{+L} \varepsilon_{ij}(s)q_i q_j ds, \qquad (1)$$

where $\varepsilon_q(s) = \varepsilon_{ij}(s)q_i q_j$ is the normal strain in the direction of the scattering vector $\boldsymbol{q}$ at location $s$ on the line along $L_1$, the direction of the secondary beam toward the neutron detector, while $\rho$ can be called the "measurement position".

It is assumed here that (i) the spatial resolution of the secondary beam is high enough so that proper volume averaging (with a certain secondary beam width $w$) can be approximated accurately with the one-dimensional integral and (ii) absorption in the sample characterised by the neutron attenuation length $d_{1/2}$ with can be neglected. In the practical terms for the axisymmetric sample of radius $R$, these two conditions can formulated as (i) $w \ll R$ and (ii) $d_{1/2} \gg R$, correspondingly.

While the full detailed derivation of the integral is given in Appendix 1, the resultant equation

$$\langle \varepsilon_q(\rho) \rangle = \frac{1}{2\sqrt{R^2-\rho^2}}\int_\rho^R A(r) \frac{dr}{\sqrt{r^2-\rho^2}} + \frac{1}{2L}\cos(2\theta)\int_\rho^R B(r) \frac{2\rho^2-r^2}{r^2} \frac{dr}{\sqrt{r^2-\rho^2}} \qquad (2)$$

has the form of an integral transform with a particular integral kernel. Although, this is a matter of serious theoretical consideration, in the current study the focus is on the feasibility of the experimental setup and measurement procedure that can provide a meaningful $\langle \varepsilon_q(\rho) \rangle$ dataset.

### 3. Experimental: neutron diffraction tomography vs neutron diffraction scanning

#### 3. 1. Material

We consider a sample made from A356 Aluminium alloy with composition (% by weight): 6.6 Si, 0.4 Mg, 0.05 Fe, 0.18 Ti, 0.019 Sr with Cu, Mn, Zn all less than 0.01. From an initial sand-casting, a bar of dimensions 25×25×140 mm$^3$ was cut and solution heat-treated at 540°C for 6 hours followed by a room-temperature water quench and ageing at 170°C for 6 hours. A 14.8 mm diameter cylinder was then machined out of the central part of the bar. This cylinder was further sliced into 3 mm thick disks to approximate plane-stress conditions. A stack of 5 discs was used in strain measurements to improve data acquisition and grain statistics, along with an assumption that the stress state in all disks were identical.

The microstructure, described previously [14], is highly uninform, comprising coarse dendrite colonies (grains) of the order of 0.8 mm in diameter with a secondary dendrite arm spacing of ~60 μm. The inter-dendritic eutectic Si inclusions were of a volume-equivalent diameter ~3 μm and an aspect ratio of ~1.6 together with age-hardening nano-precipitates of MgSi. The volume fraction of the Si inclusions is 0.062. A rather coarse microstructure of the Al matrix could have posed some technical problem due to poor grain statistics in the neutron diffraction experiment, but utilising the cylindrical symmetry of the sample, this problem was effectively eliminated by applying a continuous rotation of the sample during measurements and averaging the symmetrical positions strain values.

### 3.2. Neuron diffraction strain scanning

Neutron residual stress scanning was carried out on the KOWARI neutron diffractometer at the ANSTO OPAL research reactor [15]. Both phases, Al matrix and Si particles, were measured at 90°-geometry using wavelengths appropriate for each phase: $\lambda = 1.73$ Å for the Al(311) reflection and $\lambda = 1.58$ Å for the Si(422). Strain along three principal directions, radial, hoop and axial, was measured across the specimen diameter. The measurements in the axial direction were performed using a $2\times2\times2$ mm$^3$ gauge volume, while the gauge volume was enlarged to $2\times2\times15$ mm$^3$ for radial and hoop measurements (the larger dimension is parallel to the cylinder axis) in order to utilize the geometry of the long cylindrical sample. Due to this difference, depending on component and phase, the measurement time was changed accordingly. A typical measurement time for hoop and radial directions was 20 minutes for Si and 3 minutes for Al. These times provided strain accuracy of ~70 µstrain for the Si(422) and ~30 µstrain for the Al(311).

Due to the coarse-grained Al matrix, it was essential for measurements that sample was rotated continuously around the cylindrical axis to improve the grain statistics and to yield statistically reliable results. Additionally, and for the same reason, while measuring the axial component the cube gauge volume was moved in the axial direction by 10 mm, thus sweeping the same volume as for the hoop and radial components.

To address the critical $d_0$ problem in the residual stress analysis of a composite material, some standard pure Si powder was used to measure the strain-free lattice parameter for the same reflection Si(422). While this cannot be achieved for the Al phase, the aluminium d0 as derived indirectly through analysis of the macrostress balance conditions and separation of micro- and macro-stresses described in details in [14] on a sample of the identical material, but in case of plane-strain (long cylinder).

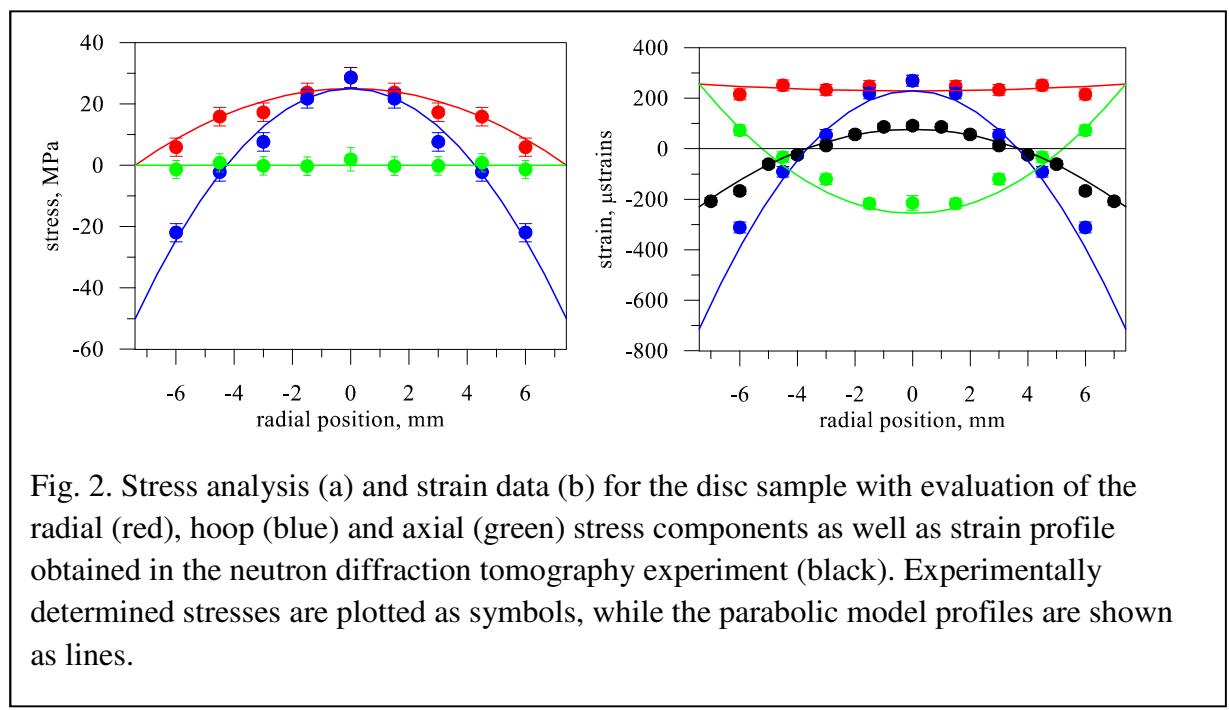

Fig. 2. Stress analysis (a) and strain data (b) for the disc sample with evaluation of the radial (red), hoop (blue) and axial (green) stress components as well as strain profile obtained in the neutron diffraction tomography experiment (black). Experimentally determined stresses are plotted as symbols, while the parabolic model profiles are shown as lines.

### 3.3. Neuron diffraction strain tomography

The neutron diffraction strain tomography experiment was carried out on the same neutron diffractometer, KOWARI. Only one phase, Al, was measured at 90°-geometry using wavelengths of λ = 2.85 Å for the Al(200) reflection using 1×15 mm² secondary slit, while primary beam was widely open to irradiate the whole sample. The sample was moved across the secondary slit in 1 mm steps to cover the whole specimen diameter. Due to a big difference in the amount of scattering material within the irradiated volume (approximately proportional to $\sqrt{R^2 - \rho^2}$, where $\rho$ is the radial position of the slice cut by the secondary slit), the measurement time was varied accordingly so that the strain accuracy was of ~30 μstrain for all points across sample. A typical measurement times for a single position $\rho$ was in the range of 10-30 minutes, so that the whole sample was measured in few hours. To eliminate the effects related to partial illimination (expected to be most severe close to the edge), strain measurements were averaged over the sample centreline.

## 4. Results

### 4.1. Neuron diffraction strain scanning

While the neutron diffraction scanning technique is well established [16] and hardly requires general validation, for this particular material and a similar sample, validation has previously been accomplished through the application of the contour method [14]. However, for the given sample yet another quality check can be used through analysis of the self-consistency of the experimental results. Stress/strain profiles in such a sample can be very accurately described using a simple single-parameter parabolic model as described in Appendix B. This single-valued model was applied to the measurement from the traditional neutron diffraction experiment with the results shown in Fig. 2a. Numerically, the model profiles were obtained using $\sigma_0 = 25$ MPa. Based on the calculated chi-square-values (for example, $\chi^2 = 0.7$ for the radial stress component and $\chi^2 = 1.0$ for the hoop component; counted per degree of freedom), the model fits experimental data sufficiently well, within statistical uncertainties.

The micro-stress from inclusions was also derived from the experiment and it was shown that it was not significant in terms of its effects on the Al phase as it was uniform across the sample, hydrostatic and of small magnitude (~7 MPa). In the following neutron diffraction strain tomography experiment, the problem of micro-stress was mitigated through choosing appropriate reference $d_0$-value to account for the shift due to micro-stress, thus effectively eliminating it.

### 4.2. Neuron diffraction strain tomography

Since the applicability of the parabolic model was demonstrated, the same model can be used for analytical prediction of $\langle \varepsilon_q(\rho) \rangle$. The result of these analytical calculations is shown in Fig. 2b against the experimentally measured values. The same statistical chi-square analysis shows that agreement between the model and experimental data is somewhat less accurate, $\chi^2 = 1.8$, but mostly due to a single point in the profile at $\rho = \pm 6\ mm$. In addition to the $\langle \varepsilon_q(\rho) \rangle$ data, the experimental strain data from the neutron diffraction scanning experiment with their fits are also plotted to demonstrate similar statistical quality of the results.

## 5. Discussion

Based on the experimental results (Fig. 2), it appears that the considered method of neutron diffraction tomography works practically as far as a sample of high symmetry is involved. The quality of data seems to be adequate and comparable with the traditional strain scanning method and this can be achieved within comparable measurement time. The spatial resolution does not appear to be a real issue since resolution of 1 mm in the neutron diffraction tomography experiment is even better that 2 mm resolution for the neutron diffraction scanning experiment. However, some difficulties in generalisation of the method to different kind of systems (material, size, symmetry) are easy to foresee;

(i) In our case of an axisymmetric stress distribution with plane-stress conditions, a single strain profile across the sample can allow the reconstruction of the full stress state under certain model assumptions that allow stress state parametrisation. In this case, the analysis is reduced to a forward integral transform problem. However, in the case of more general stress distributions, techniques for the integral transform inversion are to be developed and requirements for stress reconstruction are to be formulated. Or, at least, conditions when this inversion is possible should be pointed out.

(ii) The effects of attenuation have been neglected in our analysis. In our case of a cylindrical aluminium sample of 15 mm diameter this is appropriate; the attenuation half-length of aluminium is $d_{½} \sim 70$ mm. Clearly this will not be the case in general (e.g. $d_{½} \sim 7$ mm for iron). Thus, an additional absorption factor of form of $e^{-\mu t}$ (where $d_{½} = 1/\mu$), should appear in the integral kernel, making it at least less suitable for simple analytical treatment.

(iii) The sample size is not only to be judged against the attenuation half-length, but also against available neutron beam size and its characteristics. Some non-uniformities of the neutron flux within area of the sample might cause some complications in the data analysis (i.e. uneven weighting in the average along each slice). Also, a possible gradient of the wavelength within beam is to be controlled and this can potentially set the accuracy limit for the method.

(iv) It is assumed for this experimental method that the measured strain $\langle \varepsilon_q(\rho) \rangle$ is a linear operator in the following sense; When a diffraction pattern is measured from a large volume containing strain gradients, we assume that the sum of diffraction peak shifts coming from different sub-volumes result in an algebraic average of those shifts. In more simple terms, the position of the peak obtained by averaging of many Gaussian peaks is equal to linear combination of their positions. This is only practically true for very small shifts; strictly speaking, a sum of Gaussians is not even a Gaussian. In general case, the deviation of the diffraction peak from the reference case might be severe and special peak fitting procedure might be required.

(v) In the given study, only one strain profile was measured in the simplest condition of $2\theta = 90°$ which reduces the process to the forward Abel transform (Appendix A) [17]. The task of reconstructing the strain/stress field was further simplified using a parametric formulation; through fitting the experimentally observed $\langle \varepsilon_q(\rho) \rangle$ profile using one constant ultimately results in the establishment of all stress and strain components. However, when a parametrisation of this type cannot be used (e.g. the type of profile is unknown), even in the simple case of an axisymmetric sample with an arbitrary stress field, the application of the inverse Abel transform [17] cannot result in an unambiguous reconstruction of the stress field (apart from the additive function from the null space) since the quantity that can be reconstructed is a mixture of two components, $(\varepsilon_r + \varepsilon_\varphi)$, (Appendix A, Eq. (A13)). Therefore, it may be necessary to use measurements at $2\theta \neq 90°$, since in this case the second integral involving different combination, $(\varepsilon_r - \varepsilon_\varphi)$, is involved. Thus, measurements at different scattering angles might be a required as a part of a routine process in order to resolve $\varepsilon_r$ and $\varepsilon_\varphi$, even in simple cases. Alternatively, or perhaps additionally, an approach that imposes some constrains (e.g. equilibrium) on the stress field could be attempted in a manner that was used in the neutron transmission stress tomography [3, 4].

(vi) There are more issues mostly related to the technical details and some are listed here; A correct treatment of the partial illumination effect might be required on a new level, especially for samples of arbitrary shape. In the case of $d_0$ variation it is desirable to formulate the ways of accounting for this problem correctly. In addition to the $d_0$ variation, there might be significant effects of the preferred orientation (crystallographic texture) that through modulation of the diffraction peak intensity effectively introduce additional weight function into the integration kernel. The statistical robustness of the reconstruction methods are also to be investigated. While in the given case stress distribution is continuous, the properties of the reconstruction algorithms in general case of stress field with discontinuities are to be also investigated.

Overall, despite the limitations mentioned above, in the case of the simple sample and measurement/data analysis procedures examined here, these effects apparently were successfully eliminated or reduced to the level when they are not significant enough to affect results.

## 6. Conclusion

In this study, the practical feasibility of a neutron diffraction experiment in a tomography mode (hereby neutron diffraction tomography) was demonstrated for a simple sample with axial symmetry. The strain/stress field within an aluminium alloy sample of cylindrical geometry was measured with sufficient accuracy to be treated quantitatively and validated against a more traditional neutron diffraction stress scanning technique.

While this represents a simple demonstration of the new idea, it has proved feasibility and will perhaps bring to focus to and initiate deeper analysis of the technique. Particularly in terms of the typical questions surrounding all tomographic techniques related to inverse reconstruction. This may also spark some other technical exploration of the method at other neutron facilities, both on constant-wavelength neutron diffractometers and Time-of-Flight instruments. With measurement times of this method similar to standard stress scanning experiment, the extension of the method to more complex strain/stress fields is very possible and will come in next stages.

## Appendix A: Line integration

For the convenience of calculations it is helpful to use four coordinate systems (see Fig. A1a):
(1) The laboratory coordinate system $(X, Y)$ associated with direction $X$ being along the primary beam
(2) The coordinate system $(L_1, L_2)$ associated with direction of $L_1$ being along the secondary beam which is at the Bragg's angle $2\theta$ and $L_2$ being parallel to the intersecting perpendicular diameter
(3) The coordinate system $(e_q, e_p)$ associated with the scattering vector $q$
(4) The coordinate system $(e_r, e_\varphi)$ is a cylindrical coordinate system associated with the axisymmetric sample.

The geometric relationships (in terms of the transformation angles) between the coordinate systems that are used in our analysis are following (see Fig. A1b):
(A) Angle $\varphi$ is the angle between $L_2$ and vector $r$ (or, equivalently, $e_r$) for an arbitrary integration point $s$; This angle is related to the right triangle $\{s, \rho, r\}$ through

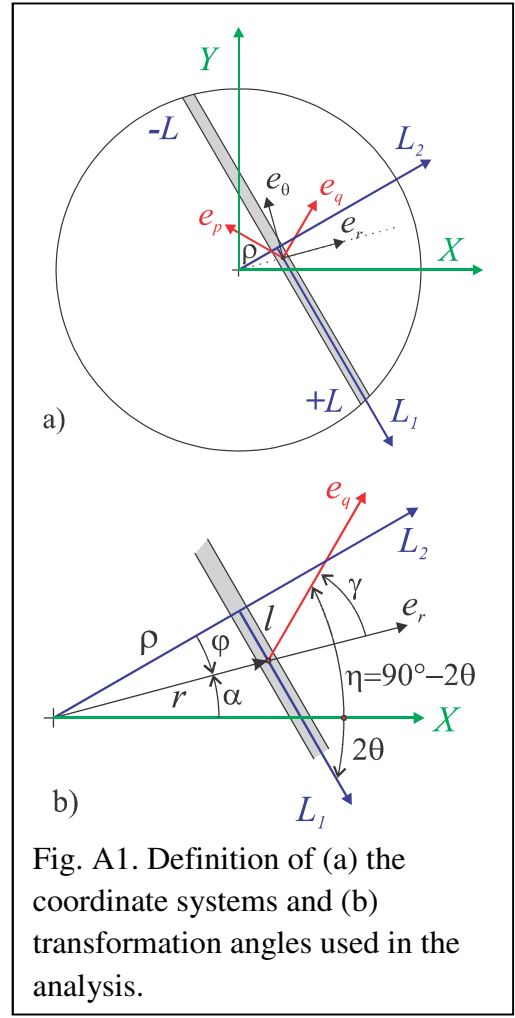

Fig. A1. Definition of (a) the coordinate systems and (b) transformation angles used in the analysis.

$$\cos(\varphi) = \rho/r;$$
$$\sin(\varphi) = s/r; \qquad (A1)$$
$$r^2 = \rho^2 + s^2 \ (R^2 = \rho^2 + L^2)$$

while the differential form being

$$ds = \frac{rdr}{s} = \frac{rdr}{\sqrt{r^2 - \rho^2}} \qquad (A2)$$

(B) For an arbitrary integration point $s$, the angle of rotation form the laboratory coordinate system to the cylindrical coordinate system is

$$\alpha = 90° - 2\theta - \varphi; \qquad (A3)$$

(C) The angle of rotation laboratory coordinate system to the scattering vector coordinate system is

$$\eta = 90° - \theta; \quad (2\eta + 2\theta = 180°) \qquad (A4)$$

(D) Finally, the angle of rotation from the cylindrical coordinate system to the scattering vector coordinate system is

$$\gamma = \eta - \alpha = 90° - \theta - 90° + 2\theta + \varphi = \theta + \varphi; \qquad (A5)$$

Using these relationship, the required integration

$$\langle \varepsilon_q(\rho) \rangle = \frac{1}{2L} \int_{-L}^{+L} \varepsilon_{ij}(l) q_i q_j ds \qquad (A6)$$

is convenient to perform in the sample cylindrical coordinates and split into two integrals

$$J_1 = \frac{1}{2L} \int_0^{+L} \varepsilon_{ij}(s) q_i q_j ds; \quad J_2 = \frac{1}{2L} \int_{-L}^{0} \varepsilon_{ij}(s) q_i q_j ds \qquad (A7)$$

Taking into account the strain transformation low
$$\varepsilon_q(\rho) = \frac{\varepsilon_r+\varepsilon_\varphi}{2} + \frac{\varepsilon_r-\varepsilon_\varphi}{2}\cos(2\gamma) = A(\rho) + B(\rho)\cos(2\gamma), \tag{A8}$$
where obviously
$$A(\rho) = \frac{\varepsilon_r+\varepsilon_\varphi}{2}; \quad B(\rho) = \frac{\varepsilon_r-\varepsilon_\varphi}{2}, \tag{A9}$$
and
$$\cos(2\gamma) = \cos(2\theta)\frac{2\rho^2-r^2}{r^2} - \sin(2\theta)\frac{2\rho\sqrt{r^2-\rho^2}}{r^2}, \tag{A10}$$
the first integral can be evaluated as
$$J_1 = \frac{1}{2\sqrt{R^2-\rho^2}}\int_\rho^R \varepsilon_q(r)\frac{rdr}{\sqrt{r^2-\rho^2}} = \tag{A11}$$
$$\frac{1}{2\sqrt{R^2-\rho^2}}\int_\rho^R A(r)\frac{rdr}{\sqrt{r^2-\rho^2}} + \frac{1}{2L}\int_\rho^R B(r)\left\{\cos(2\theta)\frac{2\rho^2-r^2}{r^2} - \sin(2\theta)\frac{2\rho\sqrt{r^2-\rho^2}}{r^2}\right\}\frac{rdr}{\sqrt{r^2-\rho^2}};$$

It can be shown that evaluation of the second integral $J_2$ gives almost identical equation to (A11) with the only difference that the term proportional to $\sin(2\theta)$ to be with the opposite sign. Therefore, the full integration (A6) results into
$$\langle\varepsilon_q(\rho)\rangle = \frac{1}{2\sqrt{R^2-\rho^2}}\left\{\int_\rho^R A(r)\frac{rdr}{\sqrt{r^2-\rho^2}} + \cos(2\theta)\int_\rho^R B(r)\frac{2\rho^2-r^2}{r^2}\frac{rdr}{\sqrt{r^2-\rho^2}}\right\} \tag{A12}$$
The two parts of integral (A12)
$$J_A(\rho) = \frac{1}{2\sqrt{R^2-\rho^2}}\int_\rho^R A(r)\frac{rdr}{\sqrt{r^2-\rho^2}} \tag{A13}$$
$$J_B(\rho) = \cos(2\theta)\frac{1}{2\sqrt{R^2-\rho^2}}\int_\rho^R B(r)\frac{2\rho^2-r^2}{r^2}\frac{rdr}{\sqrt{r^2-\rho^2}} \tag{A14}$$
are very distinct: as $J_A$ does not depend on the scattering conditions, while $J_B$ depends on them and disappears when $2\theta = 90°$. And the kernels of integral transforms are also quite different.

The first integral transformation is a variation of the Abel transform [17],
$$F(\rho) = 2\int_\rho^\infty f(r)\frac{rdr}{\sqrt{r^2-\rho^2}} \tag{A15}$$
when $f(r) = 0, r > R$.

**Appendix B: Parabolic stress model in axisymmetric plane-stress approximation**

If a parabolic law is assumed for the stress profile across axisymmetric plane-stress sample (thin disc of radius $R$), the solution is characterised by only one parameter. This, for example, can be the stress value in the center of the disk, $\sigma_0$:
$$\sigma_{rr} = \sigma_0\left(1 - \frac{r^2}{R^2}\right) \tag{B1}$$
$$\sigma_{\vartheta\vartheta} = \sigma_0\left(1 - 3\frac{r^2}{R^2}\right) \tag{B2}$$
$$\sigma_{aa} = 0 \tag{B3}$$
In this case, the Young's modulus $E$ and Poisson ration $\nu$ are also required to derive strain expressions.
$$\varepsilon_{rr} = \frac{\sigma_0}{E}\left((1-\nu) - (1-3\nu)\frac{r^2}{R^2}\right) \tag{B4}$$
$$\varepsilon_{\vartheta\vartheta} = \frac{\sigma_0}{E}\left((1-\nu) - (3-\nu)\frac{r^2}{R^2}\right) \tag{B5}$$

$$\varepsilon_{aa} = -\frac{2\nu\sigma_0}{E}(1-\nu)\left(1-2\frac{r^2}{R^2}\right) \tag{B6}$$

Equation (A12) can be analytically evaluated using (B4) and (B5) resulting in

$$\langle \varepsilon_q(r) \rangle = \frac{1}{3}\frac{\sigma_0}{E}(1-\nu)\left(1-4\frac{r^2}{R^2}\right). \tag{B7}$$